\documentclass[aps,pra,preprint,showpacs,groupedaddress,showkeys]{revtex4}

\usepackage[dvips]{graphicx}

\newcommand{\bqn}{\begin{equation}}
\newcommand{\eqn}{\end{equation}}
\newcommand{\bqna}{\begin{eqnarray}}
\newcommand{\eqna}{\end{eqnarray}}
\newcommand{\bary}{\begin{array}{clcr}}
\newcommand{\eary}{\end{array}}

\begin{document}
\title
{Two-, three-, and four-photon ionization of Mg
in the circularly and linearly polarized laser fields:\\
Comparative study using the Hartree-Fock and model potentials
}

\author{Gabriela Buica} \thanks{
Present address: Institute for Space Sciences, P.O. Box MG-23, Ro 77125,
Bucharest-M\u{a}gurele, Romania}

\affiliation
{Institute of Advanced Energy, Kyoto University,
Gokasho, Uji, Kyoto 611-0011, Japan}

\author{Takashi Nakajima}
\email[E-mail address:]{t-nakajima@iae.kyoto-u.ac.jp (Takashi Nakajima)}

\affiliation
{Institute of Advanced Energy, Kyoto University,
Gokasho, Uji, Kyoto 611-0011, Japan}

\date{\today}

%\preprint{}

\begin{abstract}
We theoretically study multiphoton ionization of Mg in the circularly as
well as the linearly polarized laser fields.
Specifically two-, three-, and four-photon ionization cross sections
from the ground and first excited states are calculated as a
function of photon energy.
Calculations are performed using the frozen-core Hartree-Fock and also
the model potential approaches and the results are compared.
We find that the model potential approach provide results as good as
or even slightly better than those by the frozen-core Hartree-Fock approach.
We also report the relative ratios of the ionization cross sections
by the circularly and linearly polarized laser fields as a function of
photon energy, which exhibit clear effects of electron correlations.
\end{abstract}


 \maketitle

%\section{}
%\subsection{}
%\subsubsection{}

% If in two-column mode, this environment will change to single-column
% format so that long equations can be displayed. Use
%\begin{widetext}
%\end{widetext}

\section{Introduction}
The interest for studying multiphoton ionization processes by using a
circularly polarized (CP) field increased in the beginning of 1970's
because of their potential of providing larger ionization cross sections
in comparison to a linearly polarized (LP) field.
Indeed, for a single-valence-electron atoms such as H and Cs, it has been
shown in Refs. \cite{gont,lamb} that the two- and three-photon ionization
cross sections by the CP field are larger than those by the LP field.
Soon after that, however, it was realized that this is not always true,
especially for multiphoton resonant and near-resonant ionization processes
in which more than a few photons are involved \cite{reiss}.
Depending on the atoms and the photon energy, cross sections by the LP
field can be larger or smaller than those by the CP field.
Reiss gave the correct interpretation of the CP and LP cross
section ratio \cite{reiss}: The ratio could be larger than one only for
processes involving up to four- or five-photon ionization.
This can be understood as a consequence of the following two reasons:
First, if the number of photons involved for ionization is more than a few,
there is more chance for the LP field than the CP field to be close to
resonance with some bound states during multiphoton absorption.
Second, since there are more ionization channels, in terms of the number
of the accessible partial waves, for the LP field than the CP field, the sum
of all these partial wave contributions could simply become larger for the
LP  field than the CP field.
We note that most of the theoretical studies performed in the above
context in those days is mainly for the single-valence-electron atoms and
there are no published results for more complex atoms in a CP field.

Time has passed since then, and because of the significant development
of theoretical methods and computer powers to calculate atomic structures,
it is now possible to calculate multiphoton ionization cross sections
for more complex atoms with reliable accuracy.
Although there are quite a few theoretical reports for the multiphoton
ionization cross sections beyond the single-valence-electron atoms
such as Xe, He, Be, Mg, and Ca
\cite{huil,saenz,chang,be_tang,spizzo,tang,chang2,gablamb2,lambro,bene},
all of them assumes the LP field.
Although theoretical data for the multiphoton ionization
cross sections for complex atoms by the CP field would provide complementary
information to those by the LP field for the purpose of understanding the
multiphoton dynamics, such data are still missing in the literature.

The purpose of this paper is to present theoretical results for the
multiphoton ionization cross sections of Mg by the CP laser field by using
the frozen-core Hartree-Fock (FCHF) and model potential (MP) approaches.
By comparing the calculated state energies and oscillator strengths,
we find that the MP approach is as good as the FCHF approach for the
oscillator strengths.
As for the state energies, the MP approach gives even better
numbers than the FCHF approach. As a result, for the purpose of calculating
multiphoton ionization cross sections of Mg, the MP approach
could give even slightly better numbers than the FCHF approach.
As additional data, we also provide the {\it ratios} of the multiphoton
ionization cross sections by the LP and CP fields, which would be very useful
from the experimental point of view, since the {\it absolute} measurement
of the multiphoton ionization cross sections is usually very difficult.
The ratio of the multiphoton ionization cross sections of Mg by the CP and
LP fields we have obtained exhibits a similar behavior with those
reported for the single-valence-electron atoms as well as rare gas atoms
\cite{gont,lamb,gui2,gui}.

\section{Theoretical approach}

The Mg atom is a {\it two-valence-electron} atom; it consists of a
closed core (the nucleus and the 10 inner-shell electrons) and
two-valence electrons.
As it is already mentioned in the literature \cite{bachau} there are
several approaches to solve the Schr\"odinger equation for
one- and two-valence-electron atom in a laser field.
Since the general computational procedure has already been presented in
Refs. \cite{chang,tang,chang2} and the specific details about the atomic
structure calculation of Mg have been reported in recent works
\cite{gablamb2,lambro}, we only briefly describe the method we employ.
The field-free one-electron Hamiltonian of Mg$^+$, $ H_{a}(r)$, is
expressed, in a.u., as:
\begin{equation}
 H_{a}(r)= -\frac{1}{2}\frac{\rm d^2}{{\rm d}r^2}
-\frac{Z}{r}+\frac{l(l+1)}{2 r^2} + V_{eff}(r),
 \label{h_1e}
\end{equation}
where $ V_{eff}(r) $ is the effective potential acting on the valence electron
of Mg$^+$. ${r}$ represents the position vector of the valence electron, $Z$
the core charge ($=2$ in our case), and $l$ the angular quantum number.
Depending on how we describe $V_{eff}(r)$, we consider two approaches
in this paper, the FCHF and the MP methods.

\subsection{One-Electron Orbitals: Frozen-Core Hartree-Fock approach}

The most widely used approach to describe the ionic core would be the
FCHF approach. In the FCHF approach the ionic core of  Mg$^{2+}$
($1 s^2 2s^2 2p^6$) is given by:
\begin{equation}
 V_{eff}(r)= V_{l}^{HF}(r) + V_l^p(r),
 \label{v_effhf}
\end{equation}
where $V_{l}^{HF}$ represents the FCHF potential (FCHFP) and
$V_l^p$ is the core-polarization potential which effectively
accounts for the interaction between the closed core and the valence
electrons \cite{chang}.
Specifically we employ the following form for the core-polarization.
$V_l^p(r)=
-{\displaystyle\frac{\alpha_s}{2 r^4}\left[1-\mbox{exp}^{-(r/{r_l})^6}\right]}$,
in which $\alpha_s$ is the static dipole polarizability of Mg$^{2+}$ and
$r_l $ the cut-off radii for the different orbital angular momenta,
$l=0,1,2,...$, etc.
For the expansion of one-electron orbitals, we employ a B-spline basis set.
Thus, solving the Schr\"odinger equation for the nonrelativistic Hamiltonian
defined in Eq. (\ref{h_1e}) with the FCHFP is now equivalent to the eigenvalue
problem.

\subsection{One-Electron Orbitals: Model potential approach}

Another simpler way to describe the ionic core is to use a MP,
$V_l^{MP}$ \cite{moccia2,aymar1,aymar2,mengali} instead of the FCHFP,
$V_{l}^{HF}$.
The advantage of the MP approach is that, we practically solve the
Schr\"odinger equation for the two-valence electrons outside the
frozen-core which is now modeled by the pseudopotential, and thus the
complexity of the problem is greatly reduced.
To start with, we employ the pseudopotential reported in \cite{preuss} to
describe the ionic core of Mg:
\begin{equation}
V_l^{MP}( r) =
- \frac{A}{r} \exp{(-\alpha r^2)}
		 + B_l \exp{(-\beta_l r^2)}
 \label{MPa},
\end{equation}
where $A  =  0.583, \alpha =  0.439, B_0 = 11.101, \beta_0 =  1.383,
B_1  =  5.220, \beta_1 = 0.995, B_{ l \geq 2} = 0$, and
$\beta_{l \geq 2} = 0$ \cite{preuss}.
Note that the core-polarization potential, $V_l^p$, is not included
in Eq. (\ref{MPa}).

In addition to the MP shown in Eq. (\ref{MPa}) (named MP$^a$),
we have obtained a new model potential with a core-polarization
term, $V_l^p$, named MP$^b$, in the following way:
We have added the core-polarization term to Eq. (\ref{MPa}) and performed
least-squares fittings for $A$ and $\alpha$ with $\alpha_s$ fixed to 0.491
\cite{moccia}, to the lowest three states of $d$ and $f$ series of Mg$^+$
simultaneously. Having fitted $A$ and $\alpha$, we proceed to perform
fittings for $B_0$ and $\beta_0$, and $B_1$ and $\beta_1$, respectively,
to the lowest three states of $s$ and $p$ series of Mg$^+$.
After these procedures, the fitted MP$^b$ parameters with the
core-polarization term are,
$A  =  0.541, \alpha =  0.561, B_0 = 11.086, \beta_0 =  1.387,
B_1  =  5.206, \beta_1 = 1.002, B_{l \geq 2} = 0$,
and $\beta_{l \geq 2} = 0$ together with the cut-off radii
 $r_0 = 1.241$, $r_1 = 1.383$, $r_2 = 1.250$,
$r_3 = 1.300$, and $r_4 = 1.100$.
We note that this form of $V_l^{MP}$ is different from the one
used in Refs. \cite{aymar1,aymar2} (named MP$^c$). In Section III, we will
make comparisons for the results obtained by FCHFP, MP$^a$, MP$^b$,
and MP$^c$.

\subsection{Two-Electron States and Ionization Cross Section}

Once the one-electron orbitals have been obtained using either FCHFP or MP,
we can construct two-electron states as we describe below:
The  field-free two-electron Hamiltonian, $ H_{a}({\bf r}_{1},{\bf
r}_{2})$, can be expressed, in a.u., as,
\begin{equation}
\label{eq:h_2e}
H_{a} ({\bf r}_{1},{\bf r}_{2})=
		\sum _{i=1}^{2}   H_a(r_i)
		+V(\mathbf{r}_{1},\mathbf{r}_{2}),
\label{HF}
\end{equation}
\noindent
where $ H_a(r_i) $ represents the one-electron Hamiltonian for the $
i^{th} $ electron as shown in Eq. (\ref{h_1e}), and
$ V(\mathbf{r}_{1},\mathbf{r}_{2})$ is a two-body interaction operator,
which comprises the static Coulomb interaction
$1/|{\bf r}_1 - {\bf r}_2| $ and the dielectronic effective
interaction \cite{chang,moccia}.
${\bf r}_1$ and ${\bf r}_2 $ are the position vectors of the two valence
electrons.
By solving the two-electron Schr\"odinger equation for the Hamiltonian
given in Eq. (\ref{HF}), two-electron states are constructed
with the configuration interaction approach \cite{chang,tang,chang2}.
Since Mg is a light atom and the LS-coupling description is known to be good,
it is sufficient to label the states by  $L$, $S$,  $J$, and $M_{J}$
which represent the total orbital angular momentum, total spin angular
momentum, total angular momentum, and its projection onto the quantization
axis, respectively.
Furthermore, if we assume that the initial state is a singlet state
($S=M_{S}=0$), $J$ and $M_{J}$, respectively, become identical with
$L$ and $M_{L}$.
Thus, the singlet states can be labeled by $L$ and $M_{L}$ only.

Once the two-electron wave functions have been obtained, the two-electron
dipole matrix elements can be calculated from state $|L  M_{L} \rangle$
to state $|L'  M_{L}' \rangle$ for both LP and CP fields.
Specifically we calculate the effective $N$-photon bound-free transition
amplitude from the ground state of Mg,
i.e., $3s^2$ $^1S^e$ ($L=0, M_L=0, S=0, M_S=0$).
The singlet-triplet transitions for the Mg atom are extremely weak and we can
safely neglect them.
Owing to the dipole selection rules for the magnetic quantum number, namely
$\Delta M_L= +1$  for the RCP field and $\Delta M_L=0$ for the LP
field, starting from the ground state with $L=M_{L}=0$, the
allowed dipole transitions by the single-photon absorption are
$\left|L,\; M_L=L \right\rangle \to
\left|L'=L+1,\; M_L'=L+1 \right\rangle $
for the CP field in comparison to
 $\left|L,\; M_L=0 \right\rangle \to
 \left|L'=L+1,\; M_L'=0 \right\rangle $ for the LP field.

Having obtained the individual two-electron dipole matrix elements we can
calculate the effective $N$-photon bound-free transition amplitude from the
initial bound state, $\left| i \right\rangle$, to the final continuum
state, $\left| f \right\rangle$, within the lowest-order perturbation theory
(LOPT). In atomic units, it reads,
\begin{eqnarray}
{\cal M}_{if}^{(N)}(\omega ) =&&
\sum_{m_1}...\sum_{m_{N-1}}
       \frac{\left\langle    f    |D_{q}| m_{N-1} \right\rangle
             \left\langle m_{N-1} |D_{q}| m_{N-2} \right\rangle}
       {E_{m_{N-1}}-E_i - (N-1)\omega}
   \nonumber \\ &&
... \times
       \frac{\left\langle m_{2} |D_{q}| m_1 \right\rangle
       \left\langle m_1 |D_{q}| i \right\rangle}
       {E_{m_1}-E_i - \omega} ,
       \label{ta}
\end{eqnarray}
\noindent
where $D_{q}$ denotes the spherical component of the dipole operator
with $q=0, \pm 1$ for the LP, RCP, and left-CP (LCP) fields,
respectively.
$E_f$, $E_g$, and $E_{m_k} $ are the state energies,
and $\omega$ represents the photon energy.
Summation is taken over all possible (both bound and continuum)
intermediate states,
$\left\{ \left| m_{k} \right\rangle \right\}$ ($k$ = $1,2,..., N-1$).
Note that Eq. (\ref{ta}) is valid for both length and velocity gauge.
In the length gauge the dipole operator is expressed as
$ D = - \textbf{E}(t) \cdot (\textbf{r}_1 +\textbf{r}_2), $
and  $D = - \textbf{A}(t) \cdot (\textbf{p}_1 +\textbf{p}_2)$ in the
velocity gauge, respectively.
$\textbf{A} (t) $ is the vector potential of the electric field vector
$\textbf{E }(t) $.
From Eq. (\ref{ta}) it should be clear that the differences of the
$N$-photon bound-free transition amplitudes by the LP and CP fields
come from the angular coefficient which is implicit in the individual
dipole matrix elements for different ${q}$, i.e.,
$\langle m_{k+1} |D_{q}| m_k \rangle$, and the accessible intermediate
as well as the final states due to the dipole selection rules as we have
illustrated above.
Finally, the generalized $N$-photon partial ionization cross section from the
ground state, $\left| g \right\rangle$, to the continuum state, $\left| c_L
\right\rangle$, is given in a.u., within LOPT, by \cite{huil},
\begin{equation}
 \sigma ^{(N)}_{gc_{L}}(\omega )
=2\pi (2\pi \alpha )^{N}\omega ^{N}
\left|{\cal M}_{gc_{L}}^{(N)}(\omega)\right|^{2} \label{cs} \; ,
\end{equation}
\noindent
where $\alpha$ represents the fine structure constant.

\section{Numerical results and discussion}

Now we present numerical results for the total and partial two-, three-, and
four-photon ionization cross sections  from the ground state of Mg by the LP
and RCP fields.
For the calculation of the individual dipole matrix elements in
Eq. (\ref{ta}), a spherical box of radius $\sim 300$ a.u. and a number of
$302$ B-spline polynomials of order $9$ and the total angular momenta
from $L=0$ up to $L=4$ are employed to represent one-electron orbitals.
In order to resolve the sharp resonance peaks appearing in the ionization
cross sections graphs, the box radius is varied from $300$ to
$320$ a.u. with a step of $1$ a.u..
We have confirmed that the calculated cross sections in both length and
velocity gauge are in very good agreement for the FCHFP approach.
As for the MP approach, however, it is well known that the correct
dipole matrix elements are given only in the length gauge
\cite{starace-kobe}, since the Hamiltonian becomes nonlocal due to
the l-dependent model potential (see Eq. (\ref{MPa})).
In other words the disagreement, if any, between the results in the length
and velocity gauge for the MP approach with l-dependent model potentials
does not imply the doubt on the reliability of the calculated results,
but rather it is a measure of the nonlocality of the l-dependent model
potential.
The same argument holds for the pseudopotential.
Therefore all the results for the MP approach reported in this paper
have been calculated in the length gauge only.

In Table I we show the comparison of calculated energies by the FCHFP,
MP$^a$, MP$^b$, and MP$^c$ \cite{aymar2}, and the experimental data for the
ionization threshold and the two-electron states for the first few low-lying
states.
All energies have been taken with respect to the energy of Mg$^{2+}$
and the experimental data have been taken from the database of
National Institute of Standards and Technology (NIST).
From Table I we notice that the MP$^a$, MP$^b$, and MP$^c$
provide more accurate energies than the FCHFP for the ionization
threshold and the first few low-lying states.
In particular the ground state energy is better described by the MP$^a$
and MP$^b$.
Of course, accuracy of the calculated energies does not always guarantee
the accuracy of the wavefunction, and we now check the accuracy of the
wavefunction in terms of the oscillator strengths.
Table II presents comparisons of the oscillator strengths for a single-photon
absorption in Mg calculated by the FCHFP, MP$^a$, MP$^b$, and MP$^c$,
and the experimental data.
The calculated values are shown for the following single-photon transitions:
$3s^2 \;^1S^e \to  3s(3-6)p \;^1P^o $,
$3s3p \;^1P^o \to  3s(4-7)s \;^1S^e $,
$3s3p \;^1P^o \to  3s(3-6)d \;^1D^e $,
$3s3d \;^1D^e \to  3s(4-7)p \;^1P^e $, and
$3s3d \;^1D^e \to  3s(4-7)f \;^1F^o $.
At first glance we see that both FCHFP and MP$^b$ give accurate values
which are comparable with the experimental data
but relatively large differences appear for the MP$^a$,
especially in case of the $3s^2 \to  3s(5-6)p $, $3s3p  \to  3s5s $,
$3s3d \to 3s(5-7)p $ transitions where the differences can reach $30\%$
for the $3s3d \to 3s5p$ and $3s7p$ transitions.
As one can easily understand, these results indicate that the inaccuracy of
the MP$^a$  most likely comes from the neglect of core-polarization.
Similarly, for the MP$^c$ some small differences appear in the
oscillator strengths for the $3s3p \to 3s(3-6)d$ and $3s3d \to 3s(5-6)p$
transitions.
From the comparisons presented in Tables I and II, it is clear that,
although both MP$^a$ and MP$^b$ provide more accurate values
for the state energies than  FCHFP, the MP$^b$ and FCHFP
provide better accuracy for the oscillator strengths.
Therefore, in what follows we present results of multiphoton ionization
cross sections obtained by the FCHFP and MP$^b$ only.

\begin{figure}
\centering
\includegraphics[width=4.5in,angle=0]{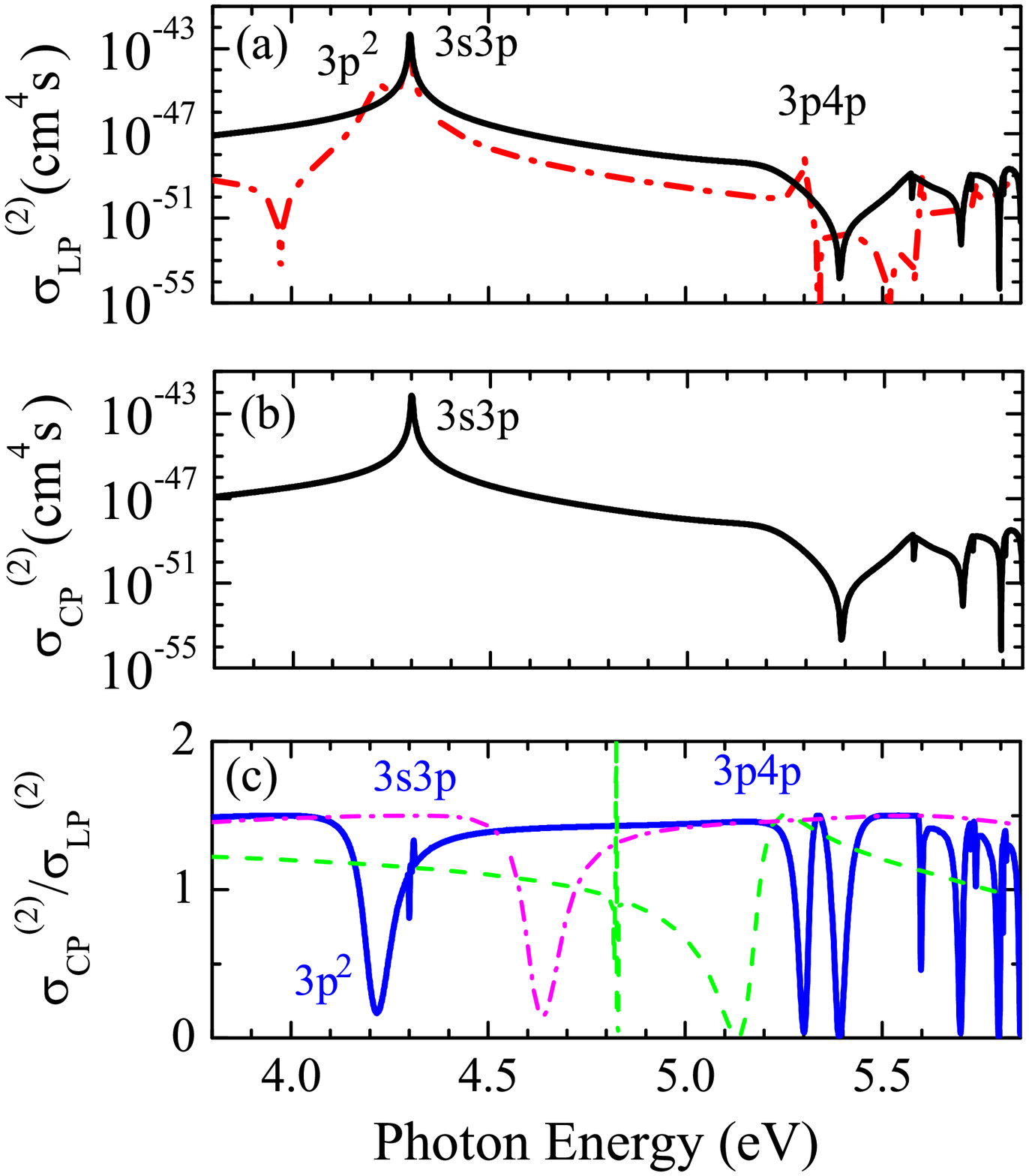}
\caption{(Color Online)
(a) Partial two-photon ionization cross sections, $ \sigma ^{(2)}_{LP} $,
by the LP field, leading to the $^{1}S$ (dashed)  and $^{1}D$ (solid)
continua, as a function of photon energy.
(b) Two-photon ionization cross section, $ \sigma ^{(2)}_{CP} $,
by the CP field.
(c) Ratio of the two-photon ionization cross sections by the CP and LP
fields,
$\sigma^{(2)}_{CP}/\sigma^{(2)}_{LP} $, as a function of photon energy.
Results for the full configuration, and simplified configurations (I) and (II)
are shown by the solid, dashed, and dot-dashed lines, respectively.}
\label{fig1}
\end{figure}

First we present numerical results for the two-, three-, and four-photon
ionization cross sections from the ground state of Mg by the LP and RCP fields
by using the FCHFP.
Figure \ref{fig1}(a) shows the partial two-photon ionization cross
sections by the LP field, leading to the $ ^{1}S ^e$ (dashed) and $ ^{1}D ^e$
continua (solid) as a function of photon energy, which are in good agreement
with those in the literature \cite{chang2,gablamb2,lambro}.
As for two-photon ionization by the CP field there is only one ionization
channel, and the ionization cross section to the $^{1}D^e$ continuum
is shown in Fig. \ref{fig1}(b).
The solid line in Fig. \ref{fig1}(c) presents the ratio of the two-photon
ionization cross sections by the CP and LP fields,
$\sigma_{CP}^{(2)}/\sigma_{LP}^{(2)}$, as a function of photon energy.
For a wide range of photon energy two-photon ionization by the CP field
turns out to be more efficient unless there is near-resonant state(s).
The ratio of the two-photon ionization cross sections is in very good
agreement with the results presented in  Refs. \cite{spizzo,lambro}.
Furthermore the maximum value of the ratio is very close to the ones
reported for the single-valence-electron atoms \cite{gont,lamb,teague,gui2}
and rare gas atoms \cite{gui}, namely $\sim 1.5$.
Although the reason why we obtain similar values seems to be simply connected
to the geometric effects (i.e., angular coefficients) starting from the
$S$ symmetry, as employed in the calculation of the dipole matrix elements
for the single-valence-electron atoms \cite{maquet}, it turned out that
the things are not so simple, as we will show later on in this paper.

\begin{figure}
\centering
\includegraphics[width=4.5in,angle=0]{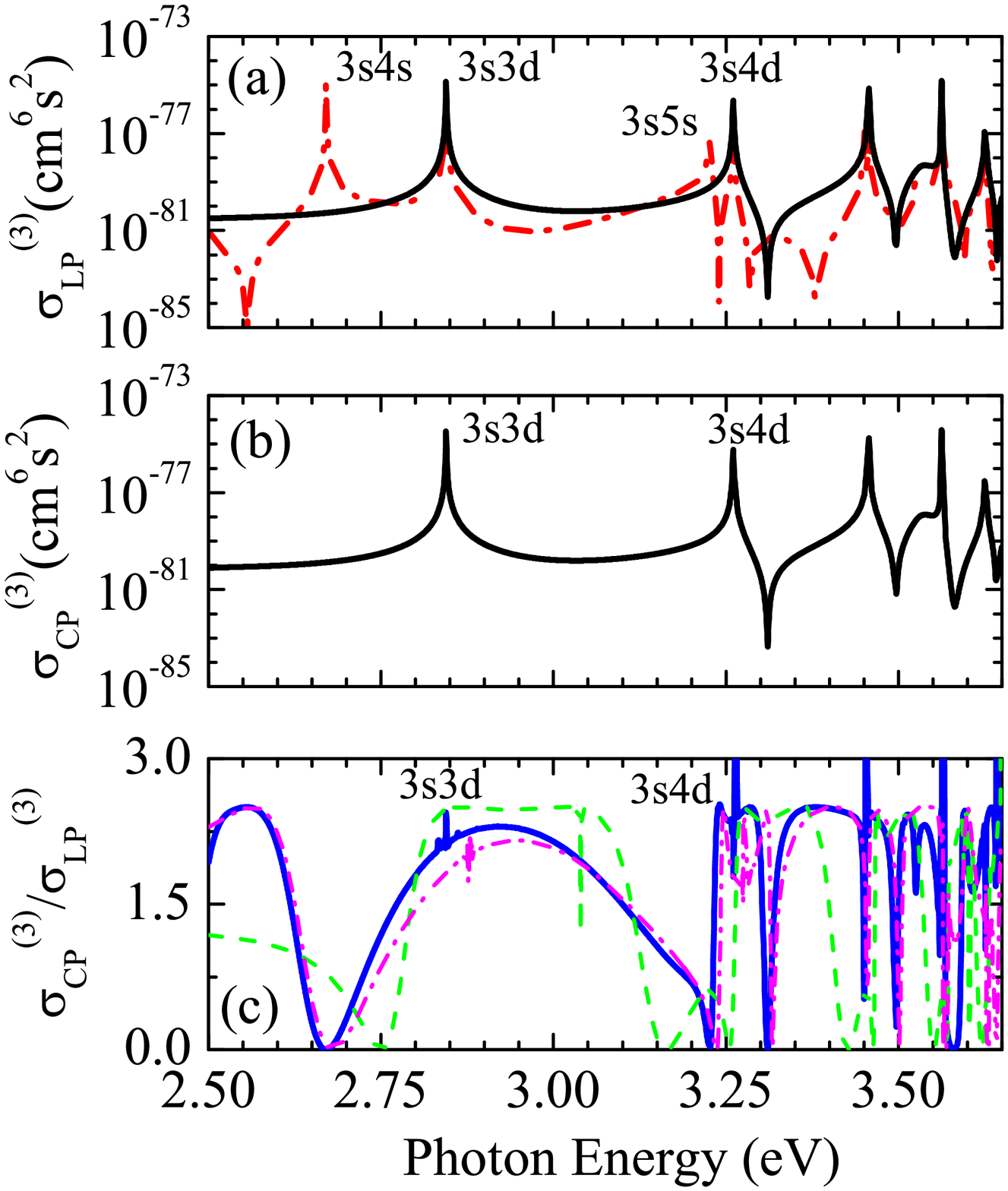}
\caption{(Color Online)(a) Partial three-photon ionization cross sections,
$ \sigma ^{(3)}_{LP} $, by LP the field, leading to the $^{1}P$ (dashed)  and
$^{1}F$ (solid) continua, as a function of photon energy.
(b) Three-photon ionization cross section $ \sigma ^{(3)}_{CP} $ by the
CP field.
(c) Ratio of the three-photon ionization cross sections by the CP and LP
fields, $\sigma^{(3)}_{CP}/\sigma^{(3)}_{LP} $, as a function of photon energy.
Results for the full configuration, and simplified configurations (I) and (II)
are shown by the solid, dashed, and dot-dashed lines, respectively.}
\label{fig2}
\end{figure}

Next we calculate the partial three-photon ionization cross sections
leading to the $ ^{1}P ^o$ (dashed) and $ ^{1}F^o $ (solid) continua.
The result is shown in Fig. \ref{fig2}(a) for the LP field.
Again our result is in good agreement with those in the literature
\cite{chang2,lambro}.
The three-photon ionization cross section by the CP field
is shown in  Fig. \ref{fig2}(b) where the accessible continuum is
$^{1}F^o$ only.
The solid line in Fig. \ref{fig2}(c) presents the ratio of the three-photon
ionization cross sections by the CP and LP fields,
$\sigma_{CP}^{(3)}/\sigma_{LP}^{(3)}$, as a function of photon energy.
Again, as in the case of two-photon ionization the maximum value
of the ratio is very close to the one reported for the
single-valence-electron atoms and rare gas atom \cite{gont,lamb,gui2,gui},
namely $\sim 2.5$.

\begin{figure}
\centering
\includegraphics[width=4.5in,angle=0]{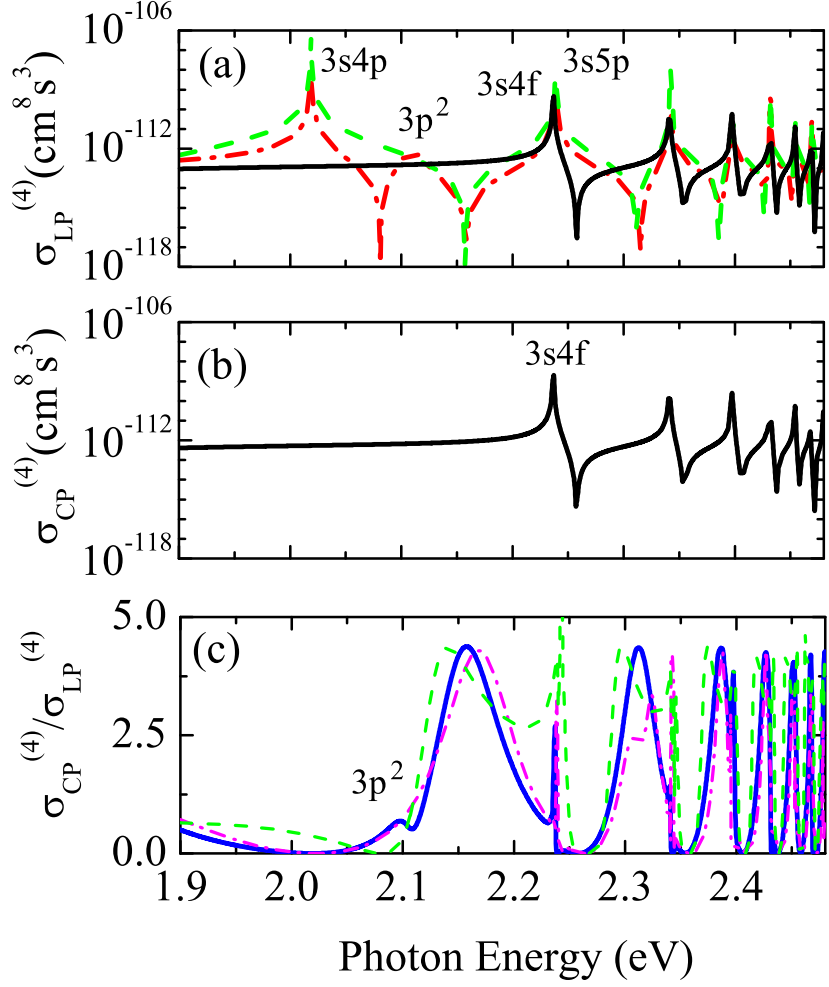}
\caption{(Color Online)
(a) Partial four-photon ionization cross sections, $ \sigma ^{(4)}_{LP} $,
by the LP field, leading to the $^{1}S$ (dot-dashed), $^{1}D$ (dashed)
and $^{1}G$ (solid) continua, as a function of photon energy.
(b) Four-photon ionization cross section, $ \sigma ^{(4)}_{CP} $, by the
CP field.
(c) Same as that in Fig. \ref{fig1}(c) but for four-photon ionization.
}
\label{fig3}
\end{figure}

Figure \ref{fig3}(a) shows the partial four-photon ionization cross
sections leading to the $ ^{1}S^e $ (dot-dashed), $ ^{1}D^e $ (dashed), and
$ ^{1}G^e $ continua (solid) as a function of photon energy for the LP field,
which is in good agreement with those in Refs. \cite{gablamb2,lambro}.
Since there are three ionization channels
for the LP field in contrast to only one ionization channel for the CP field,
 we expect that ionization by the LP field starts to be more efficient
than that by the CP field.
For the CP field the four-photon ionization cross section into the $ ^{1}G $
continuum is plotted in Fig. \ref{fig3}(b), and the ratio of the cross
sections by the CP and LP fields, $\sigma^{(4)}_{CP}/\sigma^{(4)}_{LP}$,
is shown by the solid line in Fig. \ref{fig3}(c).
The maximum value of the ratio is $\sim 4.4$ while the minimum value is
almost zero.
Figure \ref{fig3}(c) indicates that four-photon
ionization by the LP field starts to become more efficient than that
by the CP field around the photon energy of $1.9-2.1$ and $2.2-2.3$ eV.
Indeed, because of the multiphoton-character of the ionization process
and the dense energy levels of the Mg atom, more and more intermediate
states become close to resonance by the LP field compared with the CP field,
thus, contributing to the ionization efficiency.

The Mg atom is known to have strong electron correlations, and their effect
on the multiphoton ionization spectra and oscillator strengths has been
reported in, for example, Refs. \cite{chang3,chang2}.
For atoms with strong electron correlations, the values of oscillator
strengths and  ionization cross sections themselves naturally deviate
from the accurate ones if the electron correlation is not fully taken
into account.
Now we examine the effect of electron correlation in a different context:
The question we address now is whether the {\it ratios} of the multiphoton
ionization cross sections by the LP and CP fields significantly differ
with/without electron correlations taken into account.
Naively we expect that the change of the ratios would be less sensitive to
that of the cross sections themselves.
To investigate this, we have repeated the calculations for two simplified
configurations: (I) with a form of  $3snl$ ($n=3,4, ...$) only, and
 (II) $3snl$, $3p^2 \;^1P^o$, $3p(3-5)d \; ^1D^e$, and $3p(4-6)s \;^1P^o$.
The results for the configurations (I) and (II) are shown by the dashed
and dot-dashed lines in Figs. \ref{fig1}(c), \ref{fig2}(c), and \ref{fig3}(c)
for two-, three-, and four-photon ionization, respectively, with remarkable
differences. For the convenience of comparison, the ground state energy
for the simplified configuration (I) has been shifted to the accurate values
since the energy deviation turned out to be large.
Recalling that the multiphoton ionization spectra are known to be mostly
dominated by the resonant structures arising from the intermediate singly
excited bound states \cite{chang2}, we attribute these differences mainly
to the modification of the bound state wave functions, although the
differences we see for two-photon ionization (Fig. 1(c))
indicate the modification of the continuum states as well as the bound
states due to the presence of the doubly excited states $3p^{2}$ and $3p4p$.
From these comparisons it is clear that, although the maximal value
of the ratios of Mg turned out to be similar to those for
single-valence-electron atoms, the physical origin of the similarity
would not be the geometric effects, since, if so, the ratios calculated
by the simplified configurations (I) should not so much differ from others:
The angular coefficients are the same for both full and simplified
configurations.
Based on this argument, we conclude that the differences in the ratios
for the different configurations are essentially due to the
dynamic effect (electron correlation) rather than the geometric effect.

\begin{figure}
\centering
\includegraphics[width=4.5in,angle=0]{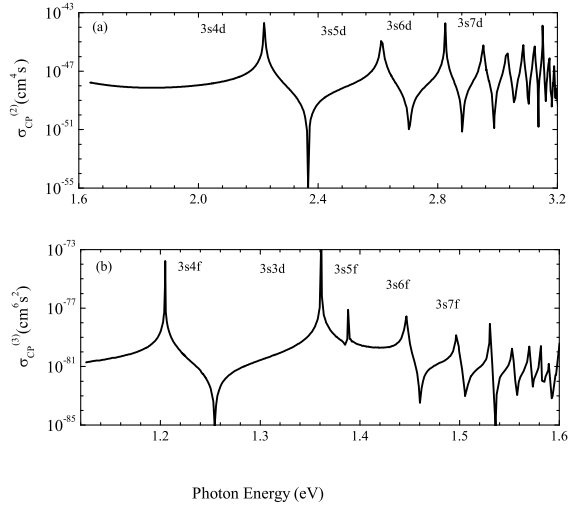}
\caption{
(a) Two-photon ionization cross section, $ \sigma ^{(2)}_{CP} $, and
(b) three-photon ionization cross section, $ \sigma ^{(3)}_{CP} $,
from the first excited state $ 3s3p \;^1P $ by the CP field.
}
\label{fig4}
\end{figure}

The results presented in Figs. \ref{fig1}-\ref{fig3} are multiphoton
ionization cross sections from the ground states.
Now, in Figs. \ref{fig4}(a) and (b), we present, for the first time,
the two- and three-photon ionization cross sections from the first
excited state $ 3s3p \;^1P^o$ of Mg by the CP laser field.
The two-photon ionization cross section from $ 3s3p \;^1P^o$ is dominated
by the single-photon resonances with the $ 3snd \;^1D^e$ ($n=4,5, ...$)
bound excited states while the three-photon ionization cross section is
dominated by the single-photon resonance due to the $ 3s3d \;^1D^e$  state
and two-photon resonances with the $ 3snf \;^1F^o$ ($n=4,5,...$) bound
excited states.

\begin{figure}
\centering
\includegraphics[width=4.5in,angle=0]{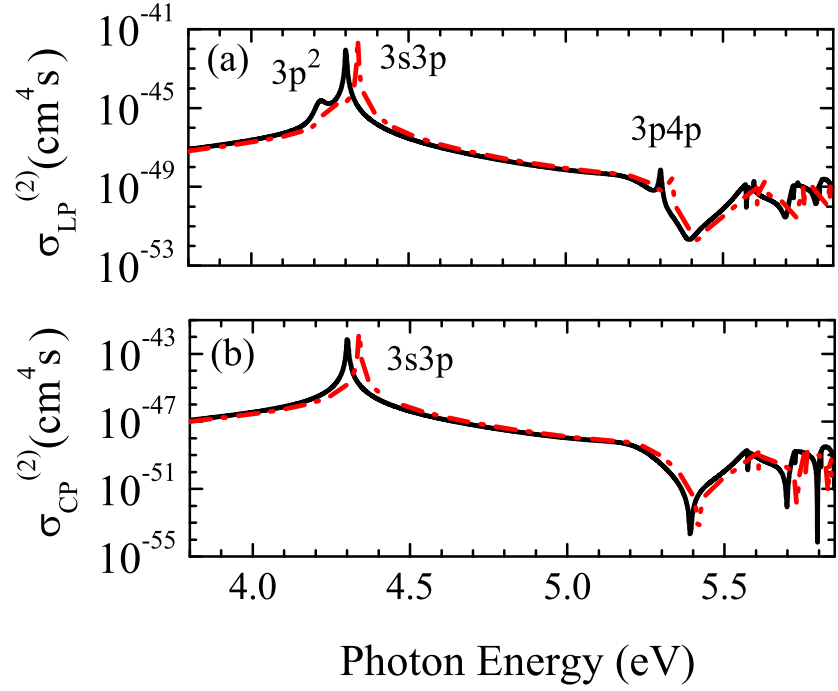}
\caption{(Color Online)
Comparison of the total two-photon ionization cross sections by the
(a) LP field, $ \sigma ^{(2)}_{LP} $, and (b) CP field,
$ \sigma ^{(2)}_{CP} $, as a function of photon energy
using the FCHFP (solid) and MP$^b$ (dot-dashed).
}
\label{fig5}
\end{figure}

\begin{figure}
\centering
\includegraphics[width=4.5in,angle=0]{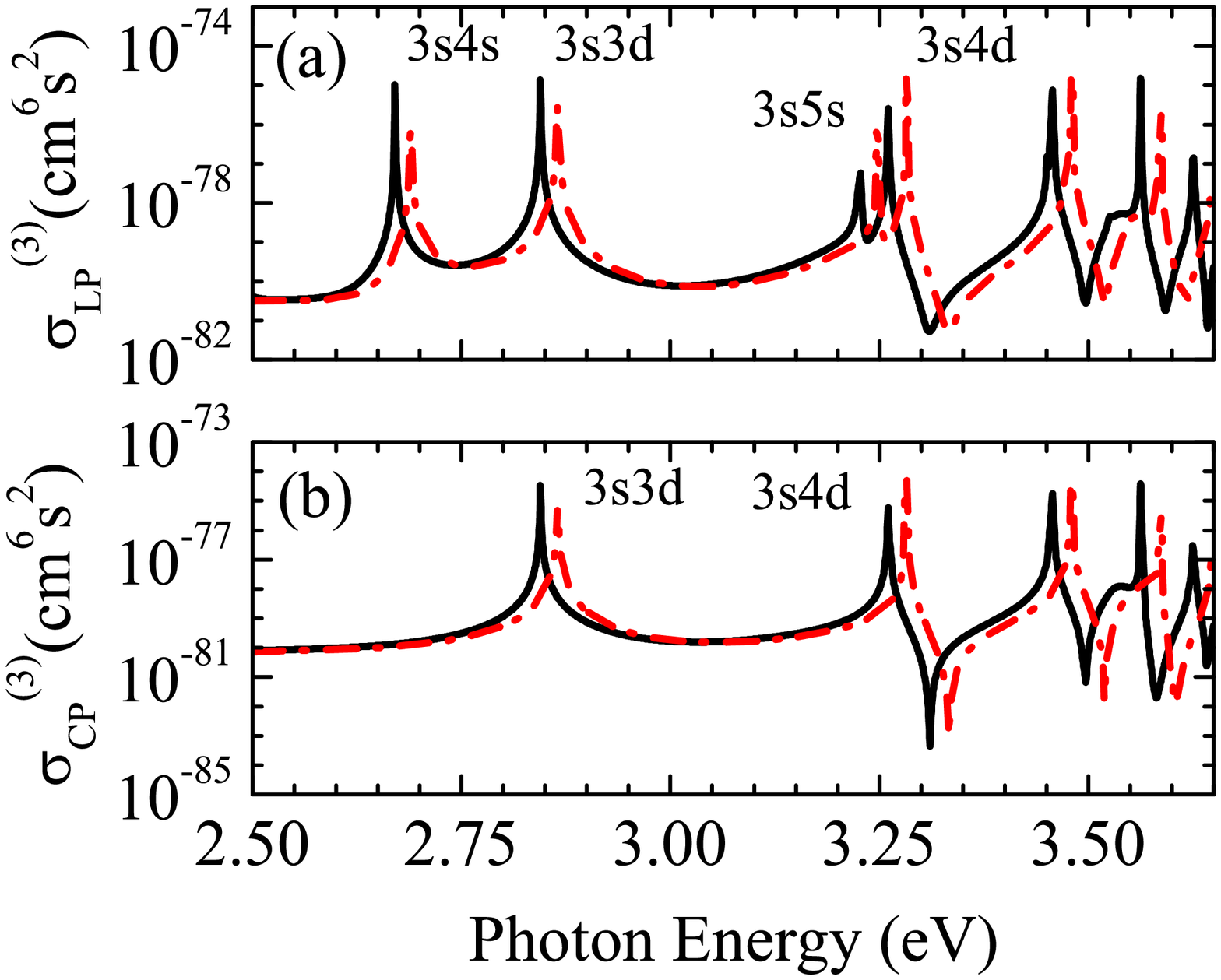}
\caption{(Color Online)
Same as those in Figs. \ref{fig5}(a) and (b) but for three-photon
ionization.}
\label{fig6}
\end{figure}

\begin{figure}
\centering
\includegraphics[width=4.5in,angle=0]{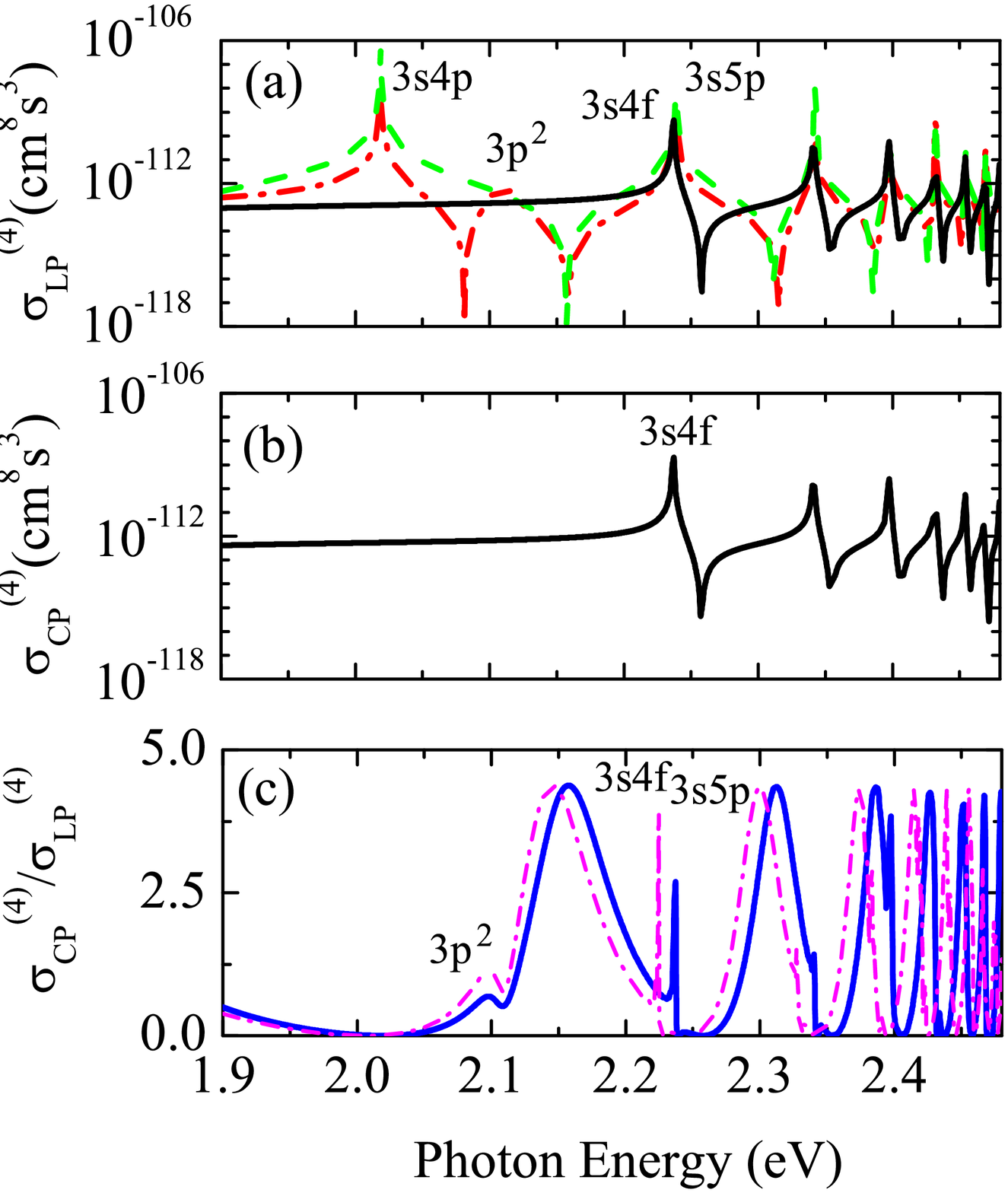}
\caption{(Color Online)
Same as those in Figs. \ref{fig5}(a) and (b) but for four-photon
ionization.}
\label{fig7}
\end{figure}

Finally, we present comparisons for the \textit{total} two-, three-,
and four-photon ionization cross sections from the ground state of Mg by
the LP and RCP fields by using the FCHFP and MP$^b$.
Comparison of the total two-photon ionization cross sections by the LP and RCP
fields using the FCHFP (solid) and MP$^b$ (dot-dashed) are
plotted in Figs. \ref{fig5} (a) and (b), respectively.
Apart from the slight energy shift, we find that the agreement is very good.
At the photon energy $\omega =4.65$ eV corresponding to the third harmonic of
a Ti-sapphire laser, the total two-photon ionization cross sections
by the LP field are 7.2 $\times 10^{-48} $ cm$^4$ s and
8.89 $\times 10^{-48} $ cm$^4$ s  calculated by
the FCHFP and MP$^b$, respectively.
Similarly, the total three- and four-photon ionization cross sections
by the LP and RCP fields using the FCHFP (solid) and MP$^b$ (dot-dashed)
are compared in Figs. \ref{fig6} and \ref{fig7}(a)-(b),
respectively.
The small energy shift which is present in all three cross section graphs
Figs. \ref{fig5}-\ref{fig7} is mainly due to the better description of
the ground state energy by the MP$^b$ than the FCHFP (see Table I).
The total three-photon ionization cross sections by the LP field
at the photon energy $\omega =3.1$ eV corresponding to the second harmonic of
a Ti-sapphire laser are 1.0 $\times 10^{-80} $ cm$^6$ s$^2$ and
1.05 $\times 10^{-80} $ cm$^6$ s$^2$ calculated by the FCHFP and MP$^b$,
respectively.
At $\omega = 2.33$ eV, Tang {\it et al.} \cite{tang} reported a value of
1.66 $\times 10^{-113} $ cm$^6$ s$^2$ which agrees
to ours of 1.86 $\times 10^{-113} $ cm$^6$ s$^2$ and
1.69 $\times 10^{-113} $ cm$^6$ s$^2$ by the FCHFP and MP$^b$,
respectively.
From these comparisons of the multiphoton ionization cross sections
presented in Figs. \ref{fig5}-\ref{fig7} and those in the literature
\cite{tang} we consider that the model potential (MP$^b$) we report
in this paper works quite well to describe the Mg atom compared with the
FCHFP with a core-polarization correction.

\section{Conclusions}

In conclusion, based on the ab-initio method, we have reported
%,for the first time,
the two-, three-, and four-photon ionization
cross sections from the ground and first excited states of Mg
by the CP field using the frozen-core Hartree-Fock method and compared
the results with those obtained by the model potential method
with a core-polarization correction. We have found that the use of
the appropriate model potential can lead to comparable or even slightly
better description for multiphoton ionization of Mg.

From the extensive calculations, we have found that depending on the photon
energy the two- and  three-photon ionization cross sections are generally
larger for the CP field  than the LP field, while starting with four-photon
ionization the result might be reversed, which qualitatively agrees with the
results for the single-valence-electron atoms and rare gas atoms
\cite{gont,lamb,gui2,gui}.
This similarity has been clearly seen in the variation of the ratios
of multiphoton ionization cross sections rations by the CP and LP fields.
The physical origin of the similarity, however, turned out to be different:
For our case at hand, it is the dynamic (electron correlation) effects
rather than the geometric effects, that determines the ratios, while
for the cases in  Refs. \cite{gont,lamb,gui2,gui} the dominant contribution
would be the geometric effects.

Related to the above, we have also studied the influence of electron
correlation in terms of the change of the {\it ratios} of multiphoton
ionization cross sections by the CP and LP fields, which we primarily
assumed to be less sensitive than the values of the cross sections
themselves, since some effects of electron correlation can be canceled out
when the ratios are taken.
Careful examinations have revealed, however, that the clear effects of
electron correlation are seen in not only the cross sections but also
in their ratios.

\section*{Acknowledgments}
G.B. acknowledges financial support from Japan Society for the
Promotion of Science.  The work by T.N. was supported by a Grant-in-Aid
for scientific research from the Ministry of Education and Science
of Japan.

\newpage
\samepage{
\begin{center}
\begin{tabular}{lp{5in}}
Table I. & Comparison of the energies for the ionization threshold
and the few bound states states of Mg.
The energies are expressed in eV with respect to the energy of Mg$^{2+}$.\\
\end{tabular}
\begin{tabular}{c c c c c c c c c c c c c c c c c c c c c c c c }
\hline \hline
     \multicolumn{1}{c}{$  $}
&&&&  \multicolumn{1}{c}{$ {\rm FCHFP}  $}
&&&& \multicolumn{1}{c}{$  {\rm MP^a}  $}
&&&&  \multicolumn{1}{c}{$ {\rm MP^b} $}
&&&&  \multicolumn{1}{c}{$ {\rm MP^c} $}
&&&&  \multicolumn{1}{c}{$ {\rm Exp}  $}
\\
\hline\hline
  \multicolumn{1}{c}{$ E_{{\rm Mg}^+} $}
&&&&  \multicolumn{1}{c}{$ -14.999555$}
&&&&  \multicolumn{1}{c}{$ -15.035553$}
&&&&  \multicolumn{1}{c}{$ -15.049135$}
&&&&  \multicolumn{1}{c}{$ -15.035241$}
&&&&  \multicolumn{1}{c}{$ -15.035266$}
 \\
\hline \hline
     \multicolumn{1}{c}{$ E_{3s3s}  $}
&&&&  \multicolumn{1}{c}{$-22.577$}
&&&&  \multicolumn{1}{c}{$-22.748	$}
&&&&  \multicolumn{1}{c}{$-22.666	$}
&&&&  \multicolumn{1}{c}{$-22.585	$}
&&&&  \multicolumn{1}{c}{$-22.681497	$}
 \\
\hline
  \multicolumn{1}{c}{$  E_{3s4s} $}
&&&&  \multicolumn{1}{c}{$ -17.236 $}
&&&&  \multicolumn{1}{c}{$ -17.301 $}
&&&&  \multicolumn{1}{c}{$ -17.288 $}
&&&&  \multicolumn{1}{c}{$ -17.264 $}
&&&&  \multicolumn{1}{c}{$ -17.2877715$}
 \\
\hline
  \multicolumn{1}{c}{$ E_{3s5s} $}
&&&&  \multicolumn{1}{c}{$-16.123 $}
&&&&  \multicolumn{1}{c}{$ -16.170$}
&&&&  \multicolumn{1}{c}{$-16.169 $}
&&&&  \multicolumn{1}{c}{$ -16.157$}
&&&&  \multicolumn{1}{c}{$ -16.1653583	$}
\\
\hline
  \multicolumn{1}{c}{$ E_{3s6s} $}
&&&&  \multicolumn{1}{c}{$ -15.677$}
&&&&  \multicolumn{1}{c}{$ -15.717$}
&&&&  \multicolumn{1}{c}{$ -15.721$}
&&&&  \multicolumn{1}{c}{$ -15.711$}
&&&&  \multicolumn{1}{c}{$ -15.715213$}
  \\
\hline \hline
     \multicolumn{1}{c}{$ E_{3s3p} $}
&&&&  \multicolumn{1}{c}{$ -18.277 $}
&&&&  \multicolumn{1}{c}{$ -18.313 $}
&&&&  \multicolumn{1}{c}{$ -18.327 $}
&&&&  \multicolumn{1}{c}{$ -18.302 $}
&&&&  \multicolumn{1}{c}{$ -18.3356944 $}
 \\
\hline
  \multicolumn{1}{c}{$ E_{3s4p}$}
&&&&  \multicolumn{1}{c}{$ -16.521 $}
&&&&  \multicolumn{1}{c}{$ -16.559 $}
&&&&  \multicolumn{1}{c}{$ -16.566 $}
&&&&  \multicolumn{1}{c}{$ -16.553$}
&&&&  \multicolumn{1}{c}{$ -16.5632827 $}
 \\
\hline
  \multicolumn{1}{c}{$  E_{3s5p}  $}
&&&&  \multicolumn{1}{c}{$ -15.860 $}
&&&&  \multicolumn{1}{c}{$ -15.898 $}
&&&&  \multicolumn{1}{c}{$ -15.904 $}
&&&&  \multicolumn{1}{c}{$ -15.894$}
&&&&  \multicolumn{1}{c}{$ -15.8987514	$}
\\
\hline
  \multicolumn{1}{c}{$  E_{3s6p} $}
&&&&  \multicolumn{1}{c}{$ -15.551	$}
&&&&  \multicolumn{1}{c}{$ -15.588	$}
&&&&  \multicolumn{1}{c}{$ -15.594	$}
&&&&  \multicolumn{1}{c}{$ -15.586$}
&&&&  \multicolumn{1}{c}{$ -15.5877425$}
  \\
\hline \hline
     \multicolumn{1}{c}{$  E_{3s3d}  $}
&&&&  \multicolumn{1}{c}{$-16.888$}
&&&&  \multicolumn{1}{c}{$-16.966	$}
&&&&  \multicolumn{1}{c}{$-16.936	$}
&&&&  \multicolumn{1}{c}{$ -16.905$}
&&&&  \multicolumn{1}{c}{$-16.9282505	$}
 \\
\hline
  \multicolumn{1}{c}{$  E_{3s4d} $}
&&&&  \multicolumn{1}{c}{$-16.056 $}
&&&&  \multicolumn{1}{c}{$-16.110 $}
&&&& \multicolumn{1}{c}{$-16.101 $}
&&&&  \multicolumn{1}{c}{$-16.082 $}
&&&&  \multicolumn{1}{c}{$-16.0936414 	$}
 \\
\hline
  \multicolumn{1}{c}{$  E_{3s5d}  $}
&&&&  \multicolumn{1}{c}{$ -15.663 $}
&&&&  \multicolumn{1}{c}{$ -15.707 $}
&&&&  \multicolumn{1}{c}{$ -15.707 $}
&&&&  \multicolumn{1}{c}{$ -15.694$}
&&&&  \multicolumn{1}{c}{$ -15.700148	$}
\\
\hline
  \multicolumn{1}{c}{$  E_{3s6d} $}
&&&&  \multicolumn{1}{c}{$ -15.451 $}
&&&&  \multicolumn{1}{c}{$ -15.491 $}
&&&&  \multicolumn{1}{c}{$ -15.494 $}
&&&&  \multicolumn{1}{c}{$ -15.484$}
&&&&  \multicolumn{1}{c}{$ -15.4875318$}
  \\
\hline \hline
     \multicolumn{1}{c}{$  E_{3s4f} $}
&&&&  \multicolumn{1}{c}{$ -15.867$}
&&&&  \multicolumn{1}{c}{$ -15.903	$}
&&&&  \multicolumn{1}{c}{$ -15.909	$}
&&&&  \multicolumn{1}{c}{$ -15.902$}
&&&&  \multicolumn{1}{c}{$ -15.9024831	$}
 \\
\hline
  \multicolumn{1}{c}{$  E_{3s5f} $}
&&&&  \multicolumn{1}{c}{$ -15.553 $}
&&&&  \multicolumn{1}{c}{$ -15.590 $}
&&&&  \multicolumn{1}{c}{$ -15.596 $}
&&&&  \multicolumn{1}{c}{$ -15.589$}
&&&&  \multicolumn{1}{c}{$ -15.5890852	$}
 \\
\hline
  \multicolumn{1}{c}{$  E_{3s6f} $}
&&&&  \multicolumn{1}{c}{$ -15.383$}
&&&&  \multicolumn{1}{c}{$ -15.419 $}
&&&&  \multicolumn{1}{c}{$ -15.426 $}
&&&&  \multicolumn{1}{c}{$ -15.419$}
&&&&  \multicolumn{1}{c}{$ -15.4190639 $}
\\
\hline
  \multicolumn{1}{c}{$  E_{3s7f} $}
&&&&  \multicolumn{1}{c}{$ -15.281 $}
&&&&  \multicolumn{1}{c}{$ -15.317 $}
&&&&  \multicolumn{1}{c}{$ -15.323 $}
&&&&  \multicolumn{1}{c}{$ -15.317$}
&&&&  \multicolumn{1}{c}{$ -15.3167415$}
  \\   \hline\hline\\
\end{tabular}
\end{center}
}

\newpage

\begin{center}
\begin{tabular}{lp{5in}}
Table II. & Comparison of the absorption oscillator strengths
(in atomic units) between states with $^1S^e$, $^1P^o $, and $^1D^e $
symmetry.
\\
\end{tabular}
\begin{tabular}{c c c c c c c c c c c c c c c}
\hline  \hline
     \multicolumn{1}{c}{$ 3s^2 \;^1S^e\rightarrow$}
&&&  \multicolumn{1}{c}{$ 3s3p \;^1P^o $}
&&&  \multicolumn{1}{c}{$ 3s4p \;^1P^o $}
&&&  \multicolumn{1}{c}{$ 3s5p \;^1P^o $}
&&&  \multicolumn{1}{c}{$ 3s6p \;^1P^o $}
 \\
\hline
     \multicolumn{1}{c}{$ {\rm FCHFP} $}
&&&  \multicolumn{1}{c}{$ 1.7659$}
&&&  \multicolumn{1}{c}{$ 0.1142$}
&&&  \multicolumn{1}{c}{$ 2.518(-2)$}
&&&  \multicolumn{1}{c}{$ 9.297(-3)$}
 \\
\hline
 \multicolumn{1}{c}{$ {\rm MP^a} $}
&&&  \multicolumn{1}{c}{$ 1.7519$}
&&&  \multicolumn{1}{c}{$ 0.1217$}
&&&  \multicolumn{1}{c}{$ 2.745(-2)$}
&&&  \multicolumn{1}{c}{$ 1.023(-2)$}
 \\
\hline
 \multicolumn{1}{c}{$ {\rm MP^b} $}
&&&  \multicolumn{1}{c}{$ 1.7677$}
&&&  \multicolumn{1}{c}{$ 0.1162$}
&&&  \multicolumn{1}{c}{$ 2.588(-2)$}
&&&  \multicolumn{1}{c}{$ 9.606(-3)$}
 \\
\hline
 \multicolumn{1}{c}{$ {\rm MP^c} $}
&&&  \multicolumn{1}{c}{$ 1.7619 $}
&&&  \multicolumn{1}{c}{$ 0.1145$}
&&&  \multicolumn{1}{c}{$ 2.535(-2)$}
&&&  \multicolumn{1}{c}{$ 9.371(-3)$}
\\
\hline
    \multicolumn{1}{c}{$ {\rm Exp}$  }
&&&  \multicolumn{1}{c}{$ 1.8$}
&&&  \multicolumn{1}{c}{$ 0.113$}
&&&  \multicolumn{1}{c}{$ 2.4 (-2) $}
&&&  \multicolumn{1}{c}{$ 9.1 (-3)$}
  \\
\hline \hline
     \multicolumn{1}{c}{$ 3s3p \;^1P^o\rightarrow$}
&&&  \multicolumn{1}{c}{$ 3s4s \;^1S^e $}
&&&  \multicolumn{1}{c}{$ 3s5s \;^1S^e $}
&&&  \multicolumn{1}{c}{$ 3s6s \;^1S^e $}
&&&  \multicolumn{1}{c}{$ 3s7s \;^1S^e $}
 \\
\hline
     \multicolumn{1}{c}{$ {\rm FCHFP} $}

&&&  \multicolumn{1}{c}{$ 0.1547$}
&&&  \multicolumn{1}{c}{$ 6.433(-3)$}
&&&  \multicolumn{1}{c}{$ 1.517(-3)$}
&&&  \multicolumn{1}{c}{$ 5.570(-4)$}
 \\
\hline
   \multicolumn{1}{c}{$ {\rm MP^a} $}
&&&  \multicolumn{1}{c}{$ 0.1553$}
&&&  \multicolumn{1}{c}{$ 5.901(-3)$}
&&&  \multicolumn{1}{c}{$ 1.403(-3)$}
&&&  \multicolumn{1}{c}{$ 5.255(-4)$}
 \\
\hline
    \multicolumn{1}{c}{$ {\rm MP^b} $}
&&&  \multicolumn{1}{c}{$ 0.1543$}
&&&  \multicolumn{1}{c}{$ 6.288(-3)$}
&&&  \multicolumn{1}{c}{$ 1.481(-3)$}
&&&  \multicolumn{1}{c}{$ 5.447(-4)$}
 \\
\hline
    \multicolumn{1}{c}{$ {\rm MP^c} $}
&&&  \multicolumn{1}{c}{$ 0.1552$}
&&&  \multicolumn{1}{c}{$ 6.453(-3)$}
&&&  \multicolumn{1}{c}{$ 1.510(-3)$}
&&&  \multicolumn{1}{c}{$ 5.493(-4)$}
 \\
\hline
     \multicolumn{1}{c}{$ {\rm Exp} $}
&&&  \multicolumn{1}{c}{$ 0.155 $}
&&&  \multicolumn{1}{c}{$ 6.3  (-3)$}
&&&  \multicolumn{1}{c}{$ 1.5  (-3)$}
&&&  \multicolumn{1}{c}{$  $}
 \\
\hline \hline
     \multicolumn{1}{c}{$ 3s3p \;^1P^o\rightarrow$}
&&&  \multicolumn{1}{c}{$ 3s3d \;^1D^e $}
&&&  \multicolumn{1}{c}{$ 3s4d \;^1D^e $}
&&&  \multicolumn{1}{c}{$ 3s5d \;^1D^e $}
&&&  \multicolumn{1}{c}{$ 3s6d \;^1D^e $}
 \\
\hline
     \multicolumn{1}{c}{$ {\rm FCHFP} $}
&&&  \multicolumn{1}{c}{$ 0.244$}
&&&  \multicolumn{1}{c}{$ 0.1076$}
&&&  \multicolumn{1}{c}{$ 0.1188$}
&&&  \multicolumn{1}{c}{$ 8.666(-2)$}
 \\
\hline
   \multicolumn{1}{c}{$ {\rm MP^a} $}
&&&  \multicolumn{1}{c}{$ 0.218$}
&&&  \multicolumn{1}{c}{$ 0.1346$}
&&&  \multicolumn{1}{c}{$ 0.1282$}
&&&  \multicolumn{1}{c}{$ 8.814(-2)$}
 \\
\hline
     \multicolumn{1}{c}{$ {\rm MP^b} $}
&&&  \multicolumn{1}{c}{$ 0.240$}
&&&  \multicolumn{1}{c}{$ 0.1025$}
&&&  \multicolumn{1}{c}{$ 0.1196$}
&&&  \multicolumn{1}{c}{$ 8.659(-2)$}
 \\
\hline
     \multicolumn{1}{c}{$ {\rm MP^c} $}
&&&  \multicolumn{1}{c}{$ 0.262$}
&&&  \multicolumn{1}{c}{$ 9.299(-2)$}
&&&  \multicolumn{1}{c}{$ 0.1010$}
&&&  \multicolumn{1}{c}{$ 8.255(-2)$}
 \\
\hline
     \multicolumn{1}{c}{$ {\rm Exp} $}
&&&  \multicolumn{1}{c}{$ 0.245 $}
&&&  \multicolumn{1}{c}{$ 0.106$}
&&&  \multicolumn{1}{c}{$ 0.121 \pm 0.01$}
&&&  \multicolumn{1}{c}{$ 8.7 (-2)$}
\\
\hline \hline
     \multicolumn{1}{c}{$ 3s3d \;^1D^e\rightarrow$}
&&&  \multicolumn{1}{c}{$ 3s4p \;^1P^o $}
&&&  \multicolumn{1}{c}{$ 3s5p \;^1P^o $}
&&&  \multicolumn{1}{c}{$ 3s6p \;^1P^o $}
&&&  \multicolumn{1}{c}{$ 3s7p \;^1P^o $}
 \\
\hline
     \multicolumn{1}{c}{$ {\rm FCHFP} $}
&&&  \multicolumn{1}{c}{$ 0.1357$}
&&&  \multicolumn{1}{c}{$ 5.886 (-3)$}
&&&  \multicolumn{1}{c}{$ 2.223 (-3)$}
&&&  \multicolumn{1}{c}{$ 1.165 (-3)$}
 \\
\hline
  \multicolumn{1}{c}{$ {\rm MP^a} $}
&&&  \multicolumn{1}{c}{$ 0.1351$}
&&&  \multicolumn{1}{c}{$ 7.494 (-3)$}
&&&  \multicolumn{1}{c}{$ 2.835 (-3)$}
&&&  \multicolumn{1}{c}{$ 1.479 (-3)$}
 \\
\hline
 \multicolumn{1}{c}{$ {\rm MP^b} $}
&&&  \multicolumn{1}{c}{$ 0.1352$}
&&&  \multicolumn{1}{c}{$ 5.963 (-3)$}
&&&  \multicolumn{1}{c}{$ 2.253 (-3)$}
&&&  \multicolumn{1}{c}{$ 1.180 (-3)$}
\\
\hline
 \multicolumn{1}{c}{$ {\rm MP^c} $}
&&&  \multicolumn{1}{c}{$ 0.1361$}
&&&  \multicolumn{1}{c}{$ 5.314 (-3)$}
&&&  \multicolumn{1}{c}{$ 2.005 (-3)$}
&&&  \multicolumn{1}{c}{$ 1.053 (-3)$}
 \\
\hline
     \multicolumn{1}{c}{$ {\rm Exp} $}
&&&  \multicolumn{1}{c}{$ 0.146  $}
&&&  \multicolumn{1}{c}{$ 5.92 (-3)$}
&&&  \multicolumn{1}{c}{$ 2.3  (-3)$}
&&&  \multicolumn{1}{c}{$ 1.1  (-3)$}
\\
\hline \hline
     \multicolumn{1}{c}{$ 3s3d \;^1D^e\rightarrow$}
&&&  \multicolumn{1}{c}{$ 3s4f \;^1F^o $}
&&&  \multicolumn{1}{c}{$ 3s5f \;^1F^o $}
&&&  \multicolumn{1}{c}{$ 3s6f \;^1F^o $}
&&&  \multicolumn{1}{c}{$ 3s7f \;^1F^o $}
 \\
\hline
     \multicolumn{1}{c}{$ {\rm FCHFP} $}
&&&  \multicolumn{1}{c}{$ 0.5150$}
&&&  \multicolumn{1}{c}{$ 0.1437$}
&&&  \multicolumn{1}{c}{$ 6.218 (-2)$}
&&&  \multicolumn{1}{c}{$ 3.321 (-2)$}
 \\
\hline
     \multicolumn{1}{c}{$ {\rm MP^a} $}
&&&  \multicolumn{1}{c}{$ 0.4742$}
&&&  \multicolumn{1}{c}{$ 0.1387$}
&&&  \multicolumn{1}{c}{$ 6.117 (-2)$}
&&&  \multicolumn{1}{c}{$ 3.3   (-2)$}
 \\
\hline
     \multicolumn{1}{c}{$ {\rm MP^b} $}
&&&  \multicolumn{1}{c}{$ 0.5082$}
&&&  \multicolumn{1}{c}{$ 0.1430$}
&&&  \multicolumn{1}{c}{$ 6.204 (-2)$}
&&&  \multicolumn{1}{c}{$ 3.319 (-2)$}
 \\
\hline
     \multicolumn{1}{c}{$ {\rm MP^c} $}
&&&  \multicolumn{1}{c}{$ 0.5336$}
&&&  \multicolumn{1}{c}{$ 0.1460$}
&&&  \multicolumn{1}{c}{$ 6.263 (-2)$}
&&&  \multicolumn{1}{c}{$ 3.33  (-2)$}

 \\
\hline
     \multicolumn{1}{c}{$ {\rm Exp}$}
&&&  \multicolumn{1}{c}{$ 0.514  $}
&&&  \multicolumn{1}{c}{$ 0.143  $}
&&&  \multicolumn{1}{c}{$ 6.2 (-2)$}
&&&  \multicolumn{1}{c}{$ 3.3 (-2)$}
 \\
\hline\hline
\end{tabular}
\end{center}

\newpage

\newpage

\newpage

\newpage

\newpage

\newpage

\newpage

\end{document}